\documentstyle[galley]{mn}
\twocolumn
\input{epsf}
\newcommand{\beq}{\begin{equation}}
\newcommand{\eeq}{\end{equation}}
\newcommand{\beqa}{\begin{eqnarray}}
\newcommand{\eeqa}{\end{eqnarray}}

\def\d{\delta}

\def\k{{\bf k}}

\def\ga{\mathrel{\mathpalette\fun >}}
\def\fun#1#2{\lower3.6pt\vbox{\baselineskip0pt\lineskip.9pt
\ialign{$\mathsurround=0pt#1\hfill##\hfil$\crcr#2\crcr\sim\crcr}}}

\title{A Fitting Formula for the Non-Linear Evolution of the Bispectrum}

\author[Scoccimarro and Couchman]{Rom\'{a}n Scoccimarro$^{1}$ and
H.M.P. Couchman$^{2}$\\ ${}^1$Institute for Advanced Study, School of
Natural Sciences, Einstein Drive, Princeton, NJ 08540, USA \\
${}^2$Department of Physics and Astronomy, McMaster University,
Hamilton, Ontario, L8S 4M1, Canada}

\pagerange{000,000}

\begin{document}
\maketitle

%
\begin{abstract}
%

We present a fitting formula for the non-linear evolution of the
bispectrum in CDM models, obtained from measurements in high
resolution numerical simulations. The formula interpolates between the
perturbative and highly non-linear regimes, and generalizes previous
results obtained for scale-free initial conditions. 

\end{abstract}


%
%
\section{Introduction}
\label{intro}
%
%

The large-scale structure of the universe observed in galaxy surveys
is thought to arise from the growth of primordial fluctuations due to
gravitational instability. Making accurate predictions of
gravitational clustering is essential if we are to compare theories with
observations and put constraints on cosmological parameters. 

Gravitational clustering is a non-linear process: when fluctuations
are sufficiently small the equations of motion can be solved
perturbatively, but at small enough scales one has resort to numerical
simulations to accurately follow the dynamics. Non-linearity of the
equations of motion implies that the mass distribution becomes
non-Gaussian, even if the initial conditions are Gaussian
(Peebles 1980, Fry 1984).

The power spectrum, or two-point correlation function, has become the
most common statistical measure of gravitational clustering. Its
importance rests on the fact that for a Gaussian random field, the
power spectrum completely characterizes its statistical
properties. Hamilton et al. (1991) suggested a useful way of thinking
about the non-linear evolution of the power spectrum, summarized by a
simple fitting formula obtained from numerical simulations. The
accuracy of this procedure has been improved (e.g. Jain, Mo \& White
1995; Peacock \& Dodds 1996) and extensively used in the literature to
predict statistical properties related to the density power spectrum.

A complete description of the non-Gaussian distribution that results
from gravitational instability, however, requires knowledge of
higher-order correlation functions. A number of useful statistics have
emerged in the last decade to become standard probes of
non-Gaussianity in large-scale structure. These include the
higher-order moments of the smoothed density field (e.g. skewness and
kurtosis; Peebles 1980; Juszkiewicz, Colombi \& Bouchet 1993;
Bernardeau 1994) and the three-point correlation function in Fourier
space, the bispectrum (Fry 1984; Scoccimarro et al. 1998, hereafter
SCFFHM) or in real space (Jing \& B\"orner 1998; Frieman \&
Gazta\~naga 1999).

A fitting formula for the non-linear evolution of one-point
higher-order moments is given by extended perturbation theory (EPT,
Colombi et al. 1997). The EPT prescription uses the perturbative
expressions for skewness, kurtosis, and so on, but with an
``effective'' spectral index that becomes a function of the variance
$\sigma^2$ of the smoothed density field. This function is obtained
numerically by fitting to measurements of the density field moments in
N-body simulations. The EPT prescription is able to fit reasonably
well the non-linear evolution of higher-order moments and thus the
one-point probability distribution function with this single function
obtained from numerical simulations.

Probing non-Gaussianity beyond one-point statistics gives precious
additional information that can be used to learn about cosmology.  The
bispectrum is the lowest-order statistic that probes the shape of
large-scale structures generated by gravitational
clustering. Measurements of the bispectrum in galaxy redshift catalogs
have been proposed as a way to constrain galaxy biasing (Fry 1994;
Hivon et al. 1995; Matarrese, Verde \& Heavens 1997) and thus break
the degeneracies present in power spectrum measurements between the
matter density of the universe $\Omega_m$ and galaxy bias. Such
measurements have recently been done in IRAS redshift surveys
(Scoccimarro et al. 2000; Feldman et al. 2000), providing constraints
on galaxy bias, $\Omega_m$ and non-Gaussian primordial
fluctuations. In addition, upcoming large redshift surveys such as 2dF
and SDSS will determine the bispectrum with unprecedented accuracy.

In this work we study the description of the non-linear evolution of
the bispectrum. A fitting formula for the bispectrum was proposed by
Scoccimarro \& Frieman (1999), hereafter SF, that interpolates between
perturbative and the highly non-linear regime, assuming scale-free
initial conditions. Since current models for structure formation are
not scale-free, these results could not be used to make realistic
predictions to be compared with observations. In this work, we extend
the SF results to provide a fitting formula for the non-linear
evolution of the bispectrum in CDM models.

\section{The Bispectrum}

We denote by $\tilde\delta$ the Fourier transform of the density
contrast $\d$ obtained from the density field $\rho=\bar{\rho}
(1+\d)$, where $\bar{\rho}$ is the mean density. The power spectrum
$P(k)$ and the bispectrum $B_{123}=B(k_1,k_2,k_3)$ are defined as

\begin{eqnarray}
\langle\tilde\delta(\k_1)\tilde\delta(\k_2)\rangle&=&
\delta_D(\k_1+\k_2)~P(k) \cr
\langle\tilde\delta(\k_1)\tilde\delta(\k_2)\tilde\delta(\k_3)\rangle&=&
\delta_D(\k_1+\k_2+\k_3)~B_{123}.
\label{spectradef}
\end{eqnarray}

\noindent In this paper, we assume Gaussian initial conditions. In this case,
the bispectrum generated by gravitational instability at large scales
is given by tree-level (second-order) perturbation theory, and reads
(Fry 1984)

\begin{equation}
B_{123}=2~F_2(\k_1,\k_2)~P_1 P_2 + {\rm cyc.},
\label{bispectredef}
\end{equation}

\noindent where $P_i \equiv P(k_i)$, and 

\begin{equation}
F_2(\k_1,\k_2)={5\over 7}+{1\over 2}{\k_1\cdot\k_2\over k_1
k_2}\left({k_1\over k_2}+ {k_2\over k_1}\right)+{2\over
7}\left({\k_1\cdot\k_2\over k_1 k_2}\right)^2.
\label{F2kernel}
\end{equation}

\noindent Note that Eq.(\ref{spectradef}) implies that the bispectrum
is defined only for closed triangles formed by the wave vectors
$\k_1,\k_2,\k_3$. It is convenient to define the reduced bispectrum
$Q_{123}$ as

\beq
Q_{123} = \frac{B_{123}}{P_1 P_2+P_2 P_3+P_3 P_1},
\label{q123}
\eeq

\noindent which takes away most of the dependence on scale and
cosmology.

SF gave a fitting formula for the non-linear evolution of the
bispectrum from scale-free initial conditions. They replaced the
kernel in Eq.~(\ref{F2kernel}) by the effective kernel

\begin{eqnarray}
F^{\rm eff}_2(\k_1,\k_2)&=&{5\over 7}~a(n,k_1)a(n,k_2)\nonumber\\
&+&{1\over 2}{\k_1\cdot\k_2\over k_1 k_2}\left({k_1\over k_2} +
{k_2\over k_1}\right)~b(n,k_1)b(n,k_2)\nonumber\\ &+&{2\over
7}\left({\k_1\cdot \k_2\over k_1 k_2}\right)^2~c(n,k_1)c(n,k_2),
\label{F2eff}
\end{eqnarray}

\noindent where the functions $a(n,k)$, $b(n,k)$, and $c(n,k)$ are
chosen to interpolate between these two regimes according to the
one-loop PT and N-body results (SCFFHM). This yields ($-2\leq n \leq
0$)

\beqa
\label{coefit1}
a(n,k) &=& \frac{1+ [0.7\ Q_3(n)]^{1/2}\ (kR_0)^{n+6}}{1 +
(kR_0)^{n+6}}, \\ & & \nonumber \\
\label{coefit2}
b(n,k) &=& \frac{1+ 0.2\ (n+3)\ (kR_0)^{n+3}}{1 + (kR_0)^{n+3.5}},
\\ & & \nonumber \\
\label{coefit3}
c(n,k) &=& \frac{1+ 4.5/[1.5+(n+3)^4]\ (kR_0)^{n+3}}{1 + (kR_0)^{n+3.5}},
\eeqa

\noindent where $R_0$ is a measure of the non-linear scale, defined
from the linear variance $\sigma^2_L(R_0) \equiv 1$ with a Gaussian
window function. The function $Q_3(n)$ is given by

\begin{equation}
Q_3(n)={(4-2^n)\over (1+2^{n+1})}.
\label{Q3def}
\end{equation}

\noindent Note that at large scales, when the functions $a=b=c=1$, we
recover the tree-level PT expression for the bispectrum; on the other
hand, at small scales $a^2=(7/10)\ Q_3$ and $b=c=0$, so
Eqs.~(\ref{bispectredef}) and~(\ref{F2eff}) give $Q_{123}=Q_3(n)$
independent of triangle configuration: this is the prediction of
hyperextended perturbation theory (HEPT), which was shown to describe
well the bispectrum (and higher order moments) in the strongly
non-linear regime (SF).

Summarizing, the non-linear bispectrum for scale-free intial
conditions is given by Eq.(\ref{bispectredef}) and
Eqs.(\ref{F2eff}-\ref{coefit3}), together with the non-linear power
spectrum obtained, for example, from the fitting formula of Peacock \&
Dodds (1996). In the next section we provide a generalization of the
bispectrum fitting formula for CDM models.

\begin{table}
\caption{Numerical Simulations Parameters}
\begin{center}
\begin{tabular}{|c|c|c|c|c|c|c|}\hline
Model & $z$ & $h(z)$ & $\Omega_m(z)$ & $\Omega_\Lambda(z)$ & $\Gamma$
& $\sigma_8(z)$ \\ \hline\hline 
SCDM & $0$ & $0.5$ & $1$ & $0$ & $0.5$ & $0.608$ \\
SCDM &$1$ & $1.4$ & $1$ & $0$ & $0.5$ & $0.304$ \\
$\tau$CDM &$0$ & $0.5$ & $1$ & $0$ & $0.21$ & $0.608$ \\ 
$\tau$CDM & $0.78$ & $1.19$ & $1$ & $0$ & $0.21$ & $0.342$ \\
OCDM & $0$ & $0.7$ & $0.3$ & $0$ & $0.21$ &$0.85$ \\
OCDM & $1$ & $1.6$ & $0.46$ & $0$ & $0.21$ & $0.575$ \\
$\Lambda$CDM & $0$ & $0.7$ & $0.3$ & $0.7$ & $0.21$ & $0.9$ \\
$\Lambda$CDM & $1$ & $1.2$ & $0.77$ & $0.23$ & $0.21$ & $0.55$ \\
\hline
\end{tabular}
\end{center}
\label{table}
\end{table}

\begin{figure*}
\epsfxsize=10truecm
\epsfysize=10truecm
\centerline{\epsffile{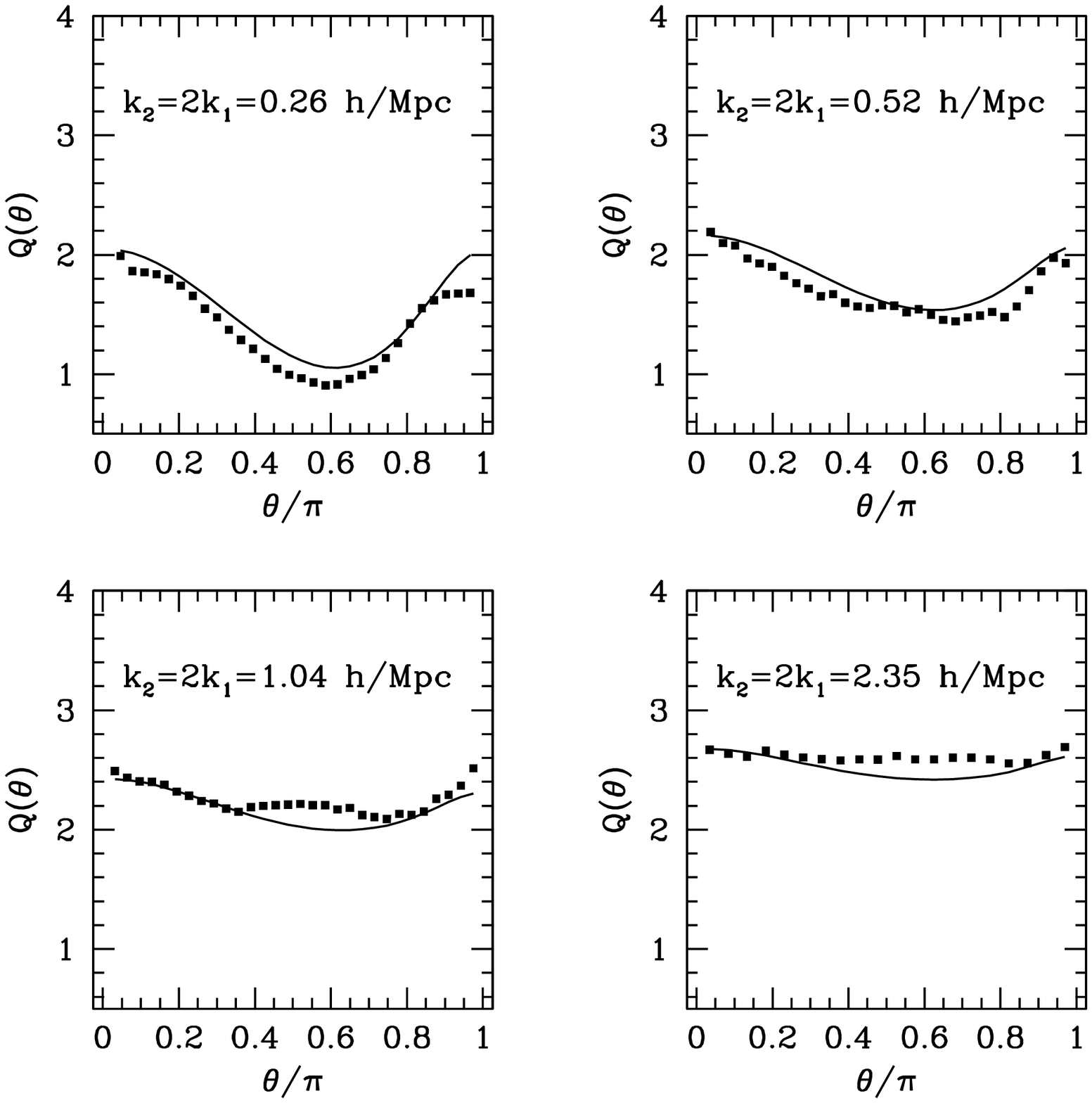}}
\caption{The bispectrum as a function of angle $\theta$ between $\k_1$
and $\k_2$ for scales as shown in the panels, for the OCDM model with
$\sigma_8=0.85$.}
\label{figOCDM}
\end{figure*}

\begin{figure*}
\epsfxsize=10truecm
\epsfysize=10truecm
\centerline{\epsffile{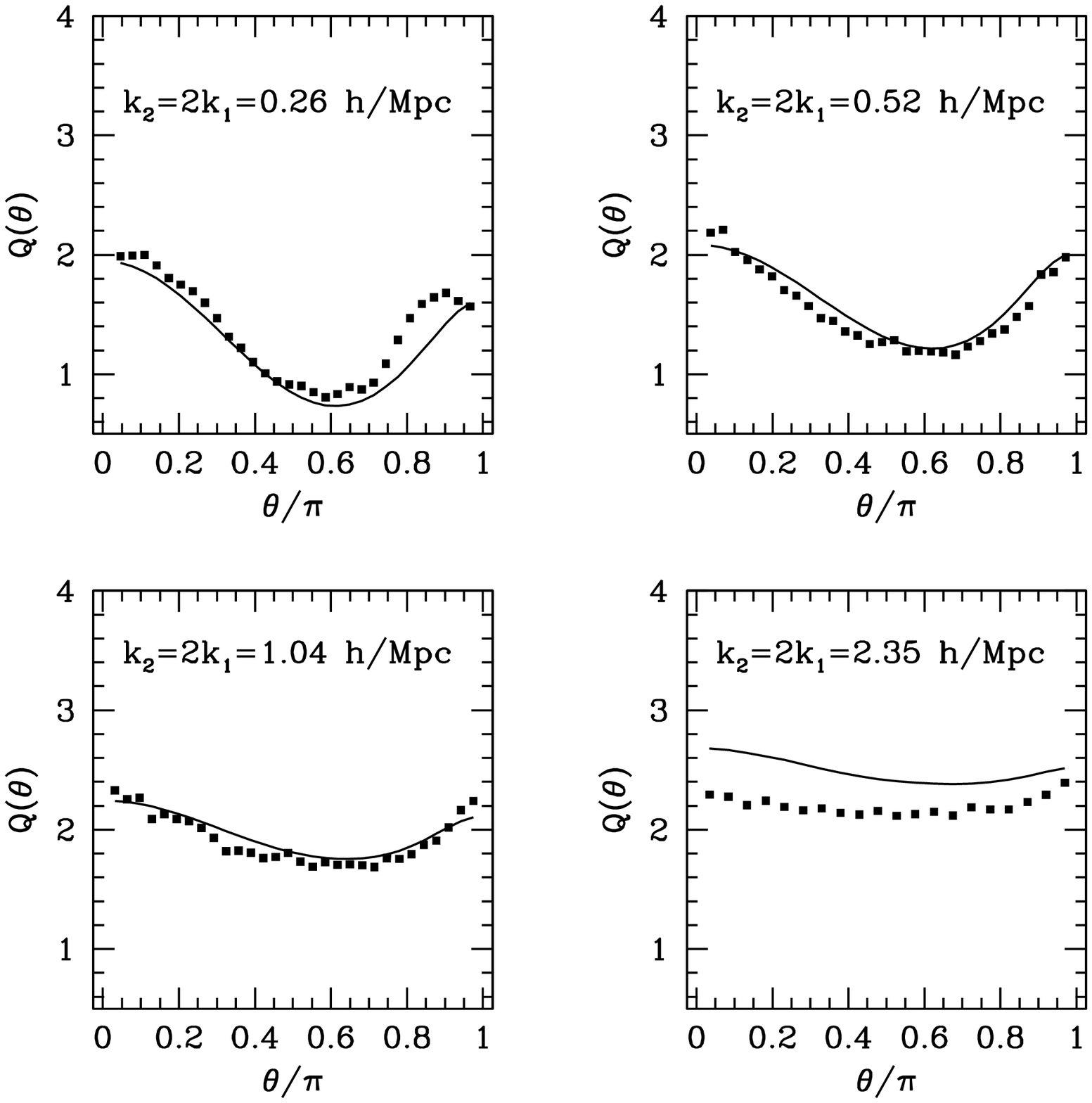}}
\caption{The bispectrum as a function of angle $\theta$ between $\k_1$
and $\k_2$ for scales as shown in the panels, for the SCDM model with
$\sigma_8=0.61$.}
\label{figSCDM}
\end{figure*}

\begin{figure*}
\epsfxsize=10truecm
\epsfysize=10truecm
\centerline{\epsffile{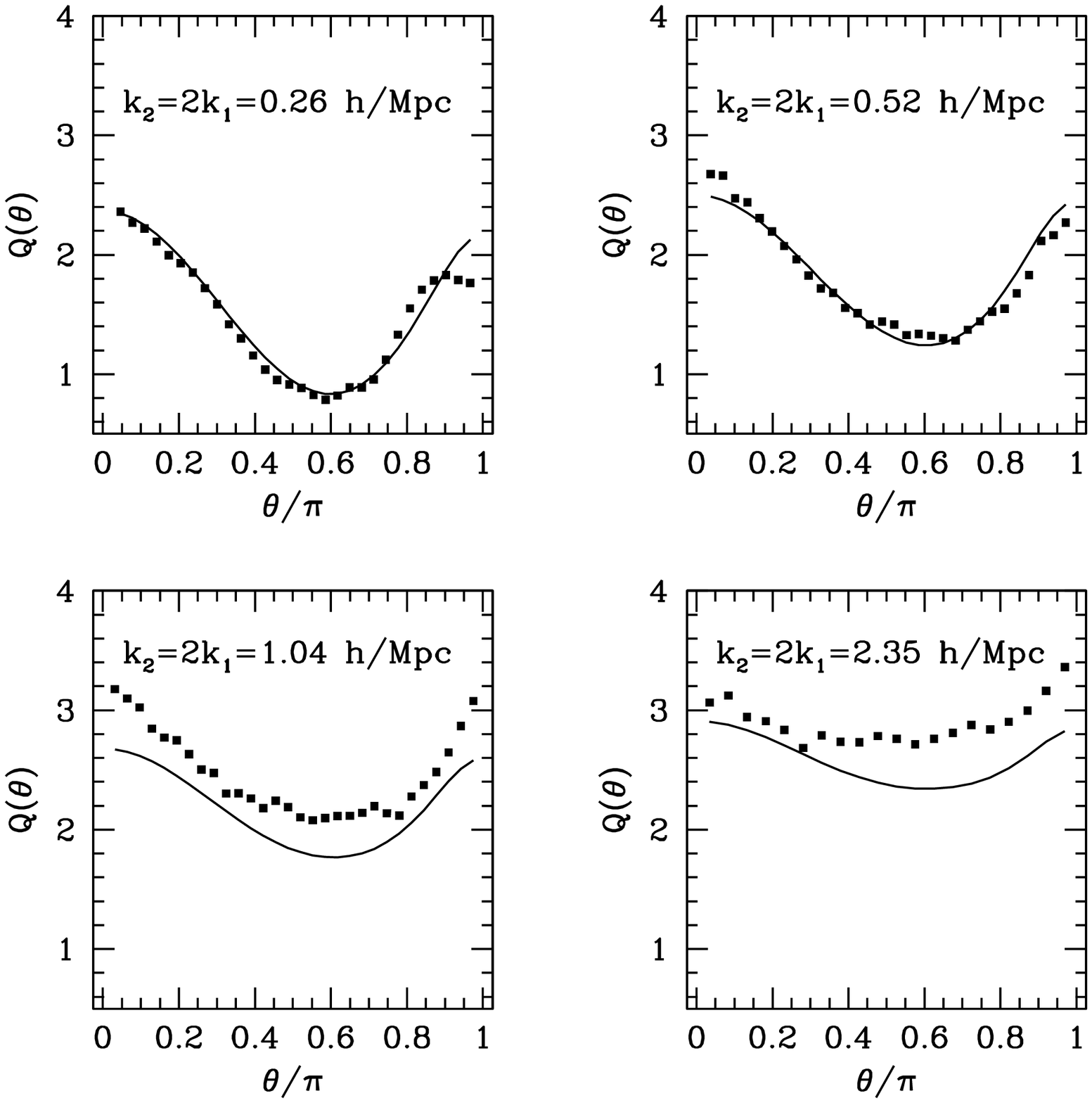}}
\caption{The bispectrum as a function of angle $\theta$ between $\k_1$
and $\k_2$ for scales as shown in the panels, for the $\Lambda$CDM
model with $\sigma_8=0.55$.}
\label{figLCDM}
\end{figure*}

\begin{figure*}
\epsfxsize=10truecm
\epsfysize=10truecm
\centerline{\epsffile{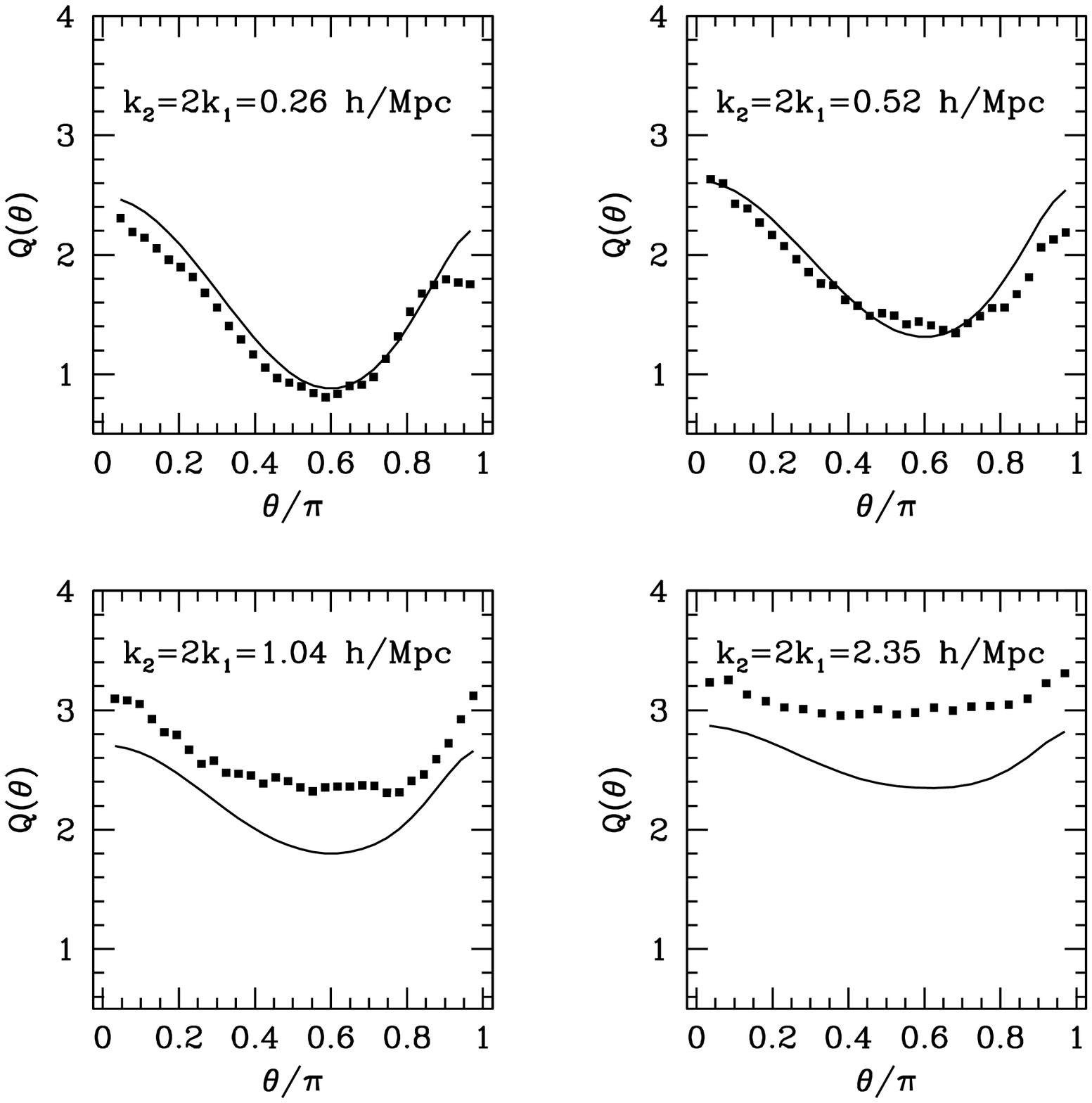}}
\caption{The bispectrum as a function of angle $\theta$ between $\k_1$
and $\k_2$ for scales as shown in the panels, for the $\tau$CDM model
with $\sigma_8=0.61$.}
\label{figTCDM}
\end{figure*}

\begin{figure*}
\epsfxsize=10truecm
\epsfysize=10truecm
\centerline{\epsffile{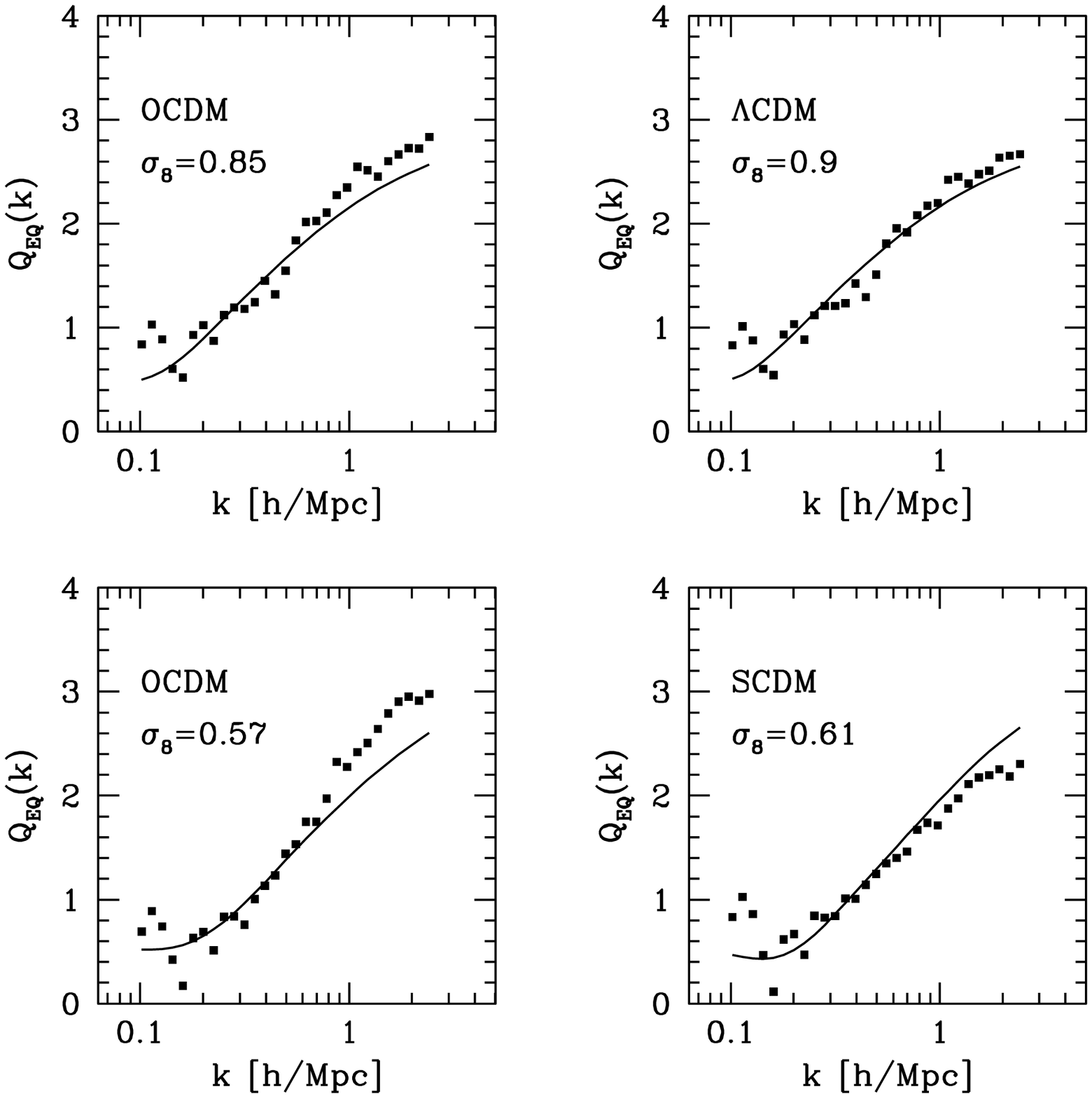}}
\caption{The bispectrum for equilateral triangles as a function of
scale. Solid lines correspond to the fitting formula, symbols to the
measurements in N-body simulations.}
\label{figeq}
\end{figure*}

%
%
\section{A Fitting Formula for CDM Models}
\label{res}
%
%

The numerical simulations used in this work correspond to
cluster-normalized cold dark matter (CDM) models run by the Virgo
collaboration (see e.g., Jenkins et al. 1998 for more details). There
are four different models, SCDM, $\tau$CDM, OCDM, and $\Lambda$CDM,
see table 1 for a description of the parameters.  These simulations
each contain $256^3$ particles in a box $240 h^{-1}$ Mpc on a
side. They were run with an adaptive P$^3$M code (Couchman, Thomas \&
Pearce, 1995; Pearce \& Couchman, 1997). There is only one realization
for each cosmological model, and all these have the same initial
phases. We have also made measurements in a set of smaller simulations
run by the Hydra Consortium (Couchman, Thomas \& Pearce, 1995), with
very similar results. The bispectrum was measured in this set of
simulations as described in SCFFHM.

Before generalizing the scale-free fitting formula (SF) reviewed in
the previous section, a number of issues related to the CDM spectrum
must be addressed. First, since the initial spectrum is no longer
scale-free, there is no unique spectral index to be used in
Eqs.~(\ref{coefit1}-\ref{coefit3}). Here we adopt the local spectral
index obtained from the linear power spectrum. Another possibility
would be to use at scale $k$ the spectral index at scale $k_L$
obtained from the HKLM mapping, $k=k_L [1+4\pi k^3 P(k)]^{1/3}$
(Hamilton et al 1991), which maps a non-linear scale to a linear scale
with the same number of enclosed pairs. Both approaches are reasonable
but can only be simple approximations to a more complex situation; in
practice the transfer of power from large to small scales depends not
only on the local spectral index but on the full shape of the power
spectrum (e.g. Jain \& Bertschinger 1994) and non-linear power at a
given wavenumber is not just a function of the linear power at another
scale. In the transition to the non-linear regime, one-loop
calculations (SCFFHM) and numerical simulations with initial power
spectra cutoff beyond some characteristic scale (Fry et al. 1993)
suggest that the local index approach is a better choice. On the other
hand, at small scales one would expect the HKLM choice to be clearly
better, since for models such as CDM where the initial power spectrum
decreases strongly with $k$, most of the power is generated by
cascading from larger scales. Despite this, approximations for
one-point higher order moments in the non-linear regime based on the
local spectral index such as EPT (Colombi et al. 1997) and HEPT (SF)
show very good agreement even for CDM spectra. Furthermore, attempts to
use the HKLM choice gave poor results (Colombi et al. 1997).

Another issue is the dependence of the fitting formula on cosmological
parameters such as $\Omega_m$ and $\Omega_\Lambda$. Perturbative
results show that the reduced bispectrum $Q$ is very weakly dependent
on cosmology at large scales, in particular, $F_2(\k_1,\k_2)$ in
Eq.~(\ref{F2kernel}) is very well approximated by its
Einstein-DeSitter value (Bouchet et al. 1995). In addition, SCFFHM
showed that as long as $\Omega_m/f^2(\Omega_m) \approx 1$, where
$f(\Omega_m) \approx \Omega_m^{0.6}$, all higher-order moments become
independent of cosmology at all orders in perturbation theory for a
fixed $\sigma_8$. For these reasons, we shall assume that the fitting
formula does not depend on cosmology; this is certainly a good
approximation within the accuracy of our measurements in
simulations. Thus, all the cosmology dependence of the bispectrum
$B_{123}$ is encoded in the cosmology dependence of the non-linear
power spectrum.

We should also note that the asymptotic behavior of the bispectrum at
small scales is not yet well tested against N-body simulations. Here
we shall follow the standard assumption of the hierarchical ansatz, in
which $Q$ becomes constant in the deeply non-linear regime, as
implicitly assumed in Eq.~(\ref{coefit1}). This behavior is consistent
with our measurements in simulations in the limited range of scales
that we can test.

To obtain a fitting formula for the bispectrum in CDM models we
proceeded as follows. First, we used the fitting formula for the
scale-free case by replacing $k R_0$ with $q= k/k_{nl}$, where $4\pi
k_{nl}^3 P_L(k_{nl})=1$, and $P_L(k)$ is the linear power spectrum at
the corresponding redshift. This provides a reasonable, but not very
accurate, fit to the simulation results, with an average deviation of
$d_{\rm avg}=40\%$ from the N-body measurements. We then perturbed
away from this fit by using two methods. We introduced six free
parameters, $a_i$, such that $q \to a_1 q$ in Eq.~(\ref{coefit1}),
$n+6 \to n+a_2$ in Eq.~(\ref{coefit1}), $0.2 \to 0.2 a_3$ in
Eq.~(\ref{coefit2}), $4.5 \to 4.5 a_4$ in Eq.~(\ref{coefit3}), $q \to
a_5 q$ in Eq.~(\ref{coefit3}), and finally $[0.7~Q_3(n)]^{1/2} \to
[0.7~Q_3(n)]^{1/2} \sigma_8^{a_6}(z)$, which quantifies deviations
from HEPT that depend on redshift. We then obtained the best fit
parameters by doing a $\chi^2$ minimization between the predicted and
measured bispectrum for all triangles, using a large grid of models
(containing about $10^6$ of them). We also looked for the best fit
using the Levenberg \& Marquardt Method (Press et al. 1992), the
solution was consistent with the grid approach but somewhat sensitive
to the initial guess for $a_i$. As a result of this procedure we get a
best fit model with parameters $a_1=0.25$, $a_2=3.5$, $a_3=2$,
$a_4=1$, $a_5=2$, $a_6=-0.2$, that is

\begin{eqnarray}
\label{coefit4}
a(n,k)&=&{1+\sigma_8^{-0.2}(z)\left[0.7~Q_3(n)\right]^{1/2}
(q/4)^{n+3.5}\over 1+(q/4)^{n+3.5}} \\
\label{coefit5}
b(n,k)&=&{1+0.4~(n+3)~q^{n+3}\over 1+q^{n+3.5}}\\ 
\label{coefit6}
c(n,k)&=&{1+4.5/\left[1.5+(n+3)^4\right](2q)^{n+3}\over 1+(2q)^{n+3.5}},
\end{eqnarray}

\noindent and $q= k/k_{nl}(z)$. The average deviation of this fitting
formula from the measurements in the simulations is about $d_{\rm
avg}=15\%$. We also found similarly good fits with $d_{\rm avg}=16\%$
for $a_1=0.25$, $a_2=4$, $a_3=1.5$, $a_4=1$, $a_5=1.5$ and $a_6=0$, so
the small deviations from HEPT implied by Eq.(\ref{coefit4}) do not
seem statistically significant.

Figures \ref{figOCDM}-\ref{figeq} illustrate the results. Although we
have measured, and used in the fitting procedure, the bispectrum for
many triangle shapes, we show for simplicity only triangles where
$k_2=2k_1$ as a function of angle $\theta$ between $\k_1$ and $\k_2$,
and equilateral triangles as a function of scale $k$. Since we only
have a single realization for each cosmological model, we cannot show
rigorous error bars, which in any case would only be part of the
story, since different triangles in a single panel are correlated:
highly so in the non-linear regime. However, we can obtain a
reasonable estimate of the errors in a single realization by scaling
the variance obtained from multiple realizations of $\Lambda$CDM lower
resolution runs (Scoccimarro et al 2000b): this gives an error in $Q$
of about $\pm 0.25$ for scales $k \ga 0.3$ h/Mpc approximately
independent of scale.

Figures~\ref{figOCDM}-\ref{figTCDM} show a comparison of the fitting
formula to the N-body measurements for triangles of different scales
(as denoted in each panel) as a function of triangle shape. The
reduced bispectrum $Q$ shows the familiar configuration dependence at
large scales (top-left panel), and then flattens out at small scales
(bottom-right panel). This behavior is well reproduced by the fitting
formula, and it is also well understood physically (SCFFHM): at large
scales the bispectrum probes the shapes of large-scale structures,
which leads to an anisotropic $Q$, whereas at small scales all
configurations become equally probable due to virialization. Note that
the flattening is more evident in higher $\sigma_8$ models, as
expected, since at a given scale these models have a stronger level of
non-linearity.

Comparison between corresponding panels in Fig.~\ref{figLCDM} and
Fig.~\ref{figTCDM} and also between the top panels in Fig.~\ref{figeq}
illustrate the lack of sensitivity of $Q$ (for a fixed level of
non-linearity, or $\sigma_8$) to cosmological parameters such as
$\Omega_m$ and $\Omega_\Lambda$, even in the non-linear regime. Note
that in some cases the agreement at the smallest scales between the
fitting formula and the simulations is not as good as at larger
scales, especially in the $\tau$CDM case; however, this does not seem
very serious given the estimated error of $\pm 0.25$ and that
different triangles are strongly correlated. More numerical work would
be necessary to assess this issue.

%
%
\section{Conclusions}
\label{conc}
%
%

We provided a fitting formula for the non-linear evolution of the
bispectrum in CDM models, which fits the measurements in N-body
simulations to an accuracy of $15\%$. This will be a useful tool for
making predictions about the non-linear behavior of non-Gaussian
statistics related to the bispectrum of the density field, and is
analagous to the role played by fitting formulas for the non-linear
evolution of the power spectrum (Hamilton et al. 1991; Jain, Mo \&
White 1995; Peacock \& Dodds 1996). Indeed, Van Waerbeke et al. (2000)
apply the results obtained here to the calculation of the skewness of
the convergence field in weak gravitational lensing surveys.

There are some issues in this work that merit further
investigation. One obvious aspect is improving the accuracy of the
fitting function; in view of the upcoming large galaxy and weak
lensing surveys it would be useful to have a more accurate prediction
for the non-linear evolution of the bispectrum. The main need in this
regard is the availability of multiple independent numerical
realizations of each cosmological model, as the bispectrum is rather
sensitive to the presence of massive clusters in the simulation volume
(Colombi et al. 1996). Another issue is the extension towards smaller
scales, and the validity of the hierarchical ansatz. Analytical models
based on dark matter halos predict that at small scales, $k \ga 3$
h/Mpc, $Q$ should increase with $k$ (Ma \& Fry 2000, Scoccimarro et
al. 2000b). Testing this prediction reliably will require the use of
large-volume, higher-resolution simulations.

\section{acknowledgments}
R.S. is supported by endowment funds from the Institute for Advanced
Study and NSF grant PHY-0070928 at IAS. H.M.P.C. Acknowledges support
from NSERC. The numerical simulations analyzed in this paper were
carried out by the Virgo Supercomputing Consortium
(http://star-www.dur.ac.uk/~frazerp/virgo/virgo.html) using computers
based at the Max Plank Institut fur Astrophysik, Garching and the
Edinburgh Parallel Computing Centre.

\end{document}